\documentclass[journal]{IEEEtran}
\usepackage{epsfig}

\ifCLASSINFOpdf
\else
\fi
\begin{document}
%
\title{On Memory Accelerated Signal Processing\\ within Software Defined Radios }
%
%
%

\author{Vincenzo Pellegrini,
        Luca Rose,
        and~Mario~Di~Dio
\thanks{V. Pellegrini and M. Di Dio are with the Dipartimento di Ingegneria dell'Informazione, Universita' di Pisa, 56122 Pisa,
Italy (e-mail: v.pellegrini@iet.unipi.it, m.didio@iet.unipi.it)}
\thanks{Manuscript submitted April 2, 2010 }}

%
%

\markboth{Submitted to arXiv Fri, April 2nd ~2010}%
{}
%



\maketitle

\begin{abstract}
Since J. Mitola's work in 1992, Software Defined Radios (SDRs) have been quite a hot topic in wireless systems research. Though many notable achievements were reported in the field, the scarcity of computational power on general purpose CPUs has always constrained their wide adoption in production environments. If conveniently applied within an SDR context, classical concepts known in computer science as \emph{space/time tradeoffs} can be extremely helpful when trying to mitigate this problem. Inspired by and building on those concepts, this paper presents a novel SDR implementation technique which we call \emph{Memory Acceleration} (MA) that makes extensive use of the memory resources available on a general purpose computing system, in order to accelerate signal computation. 
MA can provide substantial acceleration factors when applied to conventional SDRs without reducing their peculiar flexibility. As a practical proof of this, an example of MA applied in the real world to the ETSI DVB-T \cite{ETSI_DVBT} Viterbi decoder \cite{VITERBI} is provided. Actually MA is shown able to provide, when applied to such Viterbi decoder, an acceleration factor of \emph{10.4x}, with no impact on error correction performances of the decoder and by making no use of any other typical performance enhancement techniques such as low level (Assembler) programming or parallel computation, which though remain compatible with MA. Opportunity for extending the MA approach to the entire radio system, thus implementing what we call a Memory-Based Software Defined Radio (MB-SDR) is finally considered and discussed.
\end{abstract}


%
\IEEEpeerreviewmaketitle

\section{Motivation and Outline}
\label{sec:intro}
%
%
%
%
\IEEEPARstart{T}{ough} high flexibility and extremely lean and quick development cycle have always made \emph{"pure"} (i.e. based on general purpose processors without any dedicated hardware subsystem) SDRs very attractive for research, development and small-scale market deployment, low throughput per Watt compared to equivalent hardware (HW) implementations has equally kept them from accessing major markets.

It has always been a universally accepted assumption that a greater amount of generality and flexibility of the radio system had to be paid with heavy losses on power efficiency, due to the necessary usage of general purpose CPUs. Accordingly, SDR implementations up to the present date have simply aimed to replicate the classical HW-implemented signal processing chains into the software realm.

Aim of this research is to prove that, by making use of all the resources available on a general purpose computing system (i.e. not only calculus but also memory), it is possible -- at least -- to substantially reduce the power efficiency gap that exists between SDRs based on general purpose CPUs and HW implementations of the same radio system.

Implications of such results include potential for deploying flexible radio technologies on the industrial scale (due to increase of power efficiency up to levels that would make them a practical alternative to HW solutions) as well as do suggest the possibility to implement fully generic, C++ definable \emph{radio signal processing cores} being trivially derived from currently available general purpose CPUs, yet able to deliver very high signal synthesis and demodulation performances.

In order to achieve our goal, we must adapt classical concepts known in computer science under the collective denomination of \emph{space/time tradeoffs} to the signal processing realm. In previous literature, space/time tradeoffs are intended either as increasing the level of hardware (and thus often software) parallelism of a given implementation (therefore consuming more space) in order to reduce the execution time, as in \cite{STTO:1} and \cite{STTO:2} or as pre-calculating data that is already being offered by the given algorithm (or by an equivalent version of it obtained by means of some algebraic manipulation being fully peculiar to the single algorithm) into some tabular form \cite{STTO:3}. This research provides instead a convenient and rather general instrument which enables decomposition of any radio signal processing algorithm into a set of constituent parts defined in the following as \emph{segments}. A re-aggregation rule is also described for those segments, yielding a re-implementation of the given algorithm which takes conspicuous advantage from the presence of abundant memory resources.
\section{Memory Acceleration}
\subsection{System representation and taxonomy}\label{subsec:repr}
\begin{figure}[h]
	\centerline{\psfig{figure=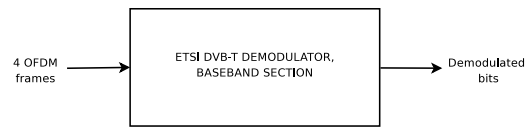,width=68.7mm} }
	\caption{DVB-T Demodulator as a Black Box}
	\label{fig:bb1}
\end{figure}

\begin{figure}[h]
	\centerline{\psfig{figure=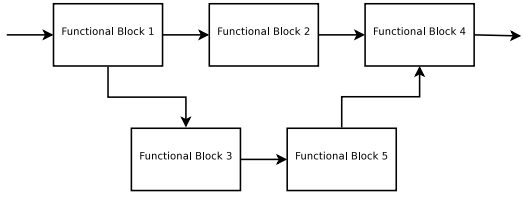,width=68.7mm} }
	\caption{Generic radio system as the web of its constituent functional blocks}
	\label{fig:chain1}
\end{figure}

We start our discussion by observing that any radio system and, more generally, any system performing some signal processing, can be represented both in a black-box fashion (figure \ref{fig:bb1}) 
and as the chain (or web) of its constituent functional blocks (figure \ref{fig:chain1}).

If we chose the black box representation of figure \ref{fig:bb1}, then we can consider our radio system (e.g. a receiver) to be fully equivalent to a mathematical function $f(...)$ which maps a certain amount of soft-valued channel symbols into the corresponding hard-valued demodulated information bits. At the receiver side, we call the minimum amount of channel symbols that can be processed independently (without requiring any other information) from the remainder of the stream the \emph{Minimum Independent Data Set} (MIDS). E.g. for ETSI DVB-T \cite{ETSI_DVBT} standard this would be 4 OFDM frames (i.e. what is called a \emph{superframe} in \cite{ETSI_DVBT}). Obviously, on the Tx path, the MIDS would be the minimum amount of information bits that the modulator will need in order to generate the atomic unit of soft-valued baseband signal defined by the radiotransmission standard. We indicate the size (number of items) of the MIDS with symbol $l$ while $A$ is the cardinality of the alphabet each datum belongs to. Domain of $f(...)$ is then defined as the set of all possible messages which can be represented within the MIDS.
We call \emph{input space} the domain of $f(...)$ and $C_{i}$ the cardinality of such space. It would then be:

\begin{equation}\label{eq:Ci}
  C_{i} = A^{l}
\end{equation}
Dually, \emph{output space} is how we name the target set of $f(...)$ while its cardinality is $C_{o}$.
Then, if we could find a convenient analytical expression for function $f(...)$, we could consider implementing our example DVB-T demodulator by programming such analytical expression into a computing system. Still, this would be a \emph{calculus-only} implementation of the system, like any classical HW or SW implementation of a radio system just is. 
I.e. such implementation would only take advantage of calculus resources being available on a general purpose computing system, disregarding the -- indeed usual and cheap -- presence of abundant memory resources.

\begin{figure}[h]
	\centerline{\psfig{figure=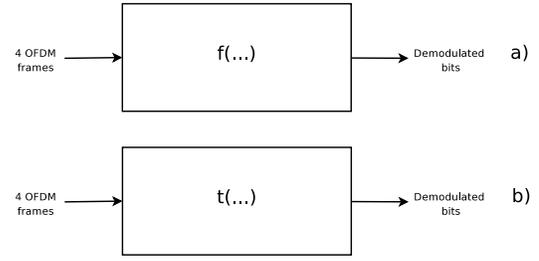,width=68.7mm} }
	\caption{Same example system being implemented through pure calculus a) as well as pure memory b)}
	\label{fig:ft}
\end{figure}

This said, it would be natural and obvious to think of replacing function $f(...)$ with table $t(...)$: a table containing, for each item of the input space, the associated output value as shown in figure \ref{fig:ft}. This would be a \emph{memory-only} implementation, would \emph{not require extraction of $f(...)$ in analytical form} but would obviously yield a $C_{i}$ being not practical for any memory technology available today or in the foreseeable future. Actually, the table $t(...)$ could be filled up by running the standard, computation-only, implementation of function $f(...)$ (that is the radio signal processing chain implementing it) over the entire input space at instantiation time. 

Such considerations suggest that the path towards optimal SDRs lies somewhere among hybrid systems which are able to use up both calculus and memory.  

We now move a step forward by abandoning the single black-box representation of our SDR system which we used as a tool to describe the fundamental rationale behind the MA idea. 
We begin the process of designing Memory Acceleration for our SW radio and, in order to do this, we go deeper into the knowledge of the SDR we wish to implement by representing it as the web of its constituent blocks, see figure \ref{fig:chain1}. 

We call this representation change the \emph{0-step} of \emph{algorithm segmentation} because at first we consider our radio system to be a single signal processing algorithm and then we break it down into its functional blocks. Obviously such  \emph{0-step} comes for free as long as any radiotransmission system is conceived since the beginning as a chain of functional blocks.

Actually, at this early stage we assume each of the functional blocks $f_{n}(...)$ to be \emph{atomic}, i.e. impossible to break in a further web of constituent algorithmic functional blocks, still, in subsequent phases of MA design, \emph{algorithm segmentation} will be developed further as it will play a key-role for the effectiveness of the MA / MB-SDR implementation strategy. Precisely, within the scope of this work, \emph{algorithm segmentation} is defined as the decomposition of a functional block $f(...)$, formerly assumed to be atomic, into a chain (or web) of constituent sub-blocks $f_{n}(...)$ which implements the same function as $f(...)$. Input spaces of sub blocks $C_{i_{n}}$ will be different from and significantly smaller than $C_{i}$, if algorithm segmentation was performed correctly. 

It must be noted that algorithm segmentation is not algorithm re-design: in particular this yields that algorithm segmentation \emph{does not change the overall computational cost of the segmented algorithm}.

\begin{table}
  \renewcommand{\arraystretch}{1.0}
  \caption{MA / MB-SDR symbols and taxonomy}
  \label{tab:parallel}
  \centering
  \begin{tabular}{c|c}
    \hline
    \emph{Symbol} & \emph{Meaning}\\
    \hline
    \hline
    $f_{n}(...)$ & Computation-only implementation of block n \\
    \hline
    $t_{m}(...)$ & Memory table m \\
    \hline
    $l$ & Number of items within the MIDS \\
    \hline
    $A$ & Cardinality of alphabet for each item of MIDS  \\
    \hline
    $C_{i_{n}}$ & Cardinality of input space of block $f_{n}(...)$\\
    \hline
    $C_{o_{n}}$ & Cardinality of output space of block $f_{n}(...)$\\
    \hline
    $W$ & Total available computational power \\
    \hline
    $W_{n}$ & Computational cost of block $f_{n}(...)$\\
    \hline
    $W_{TB}$ & Computational cost of subsystem within table boundary\\
    \hline
    $W_{m}$ & Computational cost of subsystem replaced by table m\\
    \hline
    $Wm_{m}$ & Computational cost for handling table $t_{m}(...)$\\
    \hline
    $M$ & Total size of available memory  \\
    \hline
    $M_{m}$ & Total memory footprint of table $t_{m}(...)$\\
    \hline
    $S_{m}$ & Data size of items stored in  $t_{m}(...)$\\
    \hline
    $a$ & Acceleration factor \\
    \hline
    $\eta$ & Acceleration efficiency\\
    \hline
    $\eta_{m}$ & Acceleration efficiency of table $t_{m}(...)$\\
    \hline
    I &  Overall SDR implementation merit parameter\\ 
    \hline
  \end{tabular}
\end{table}

\begin{figure}[h]
	\centerline{\psfig{figure=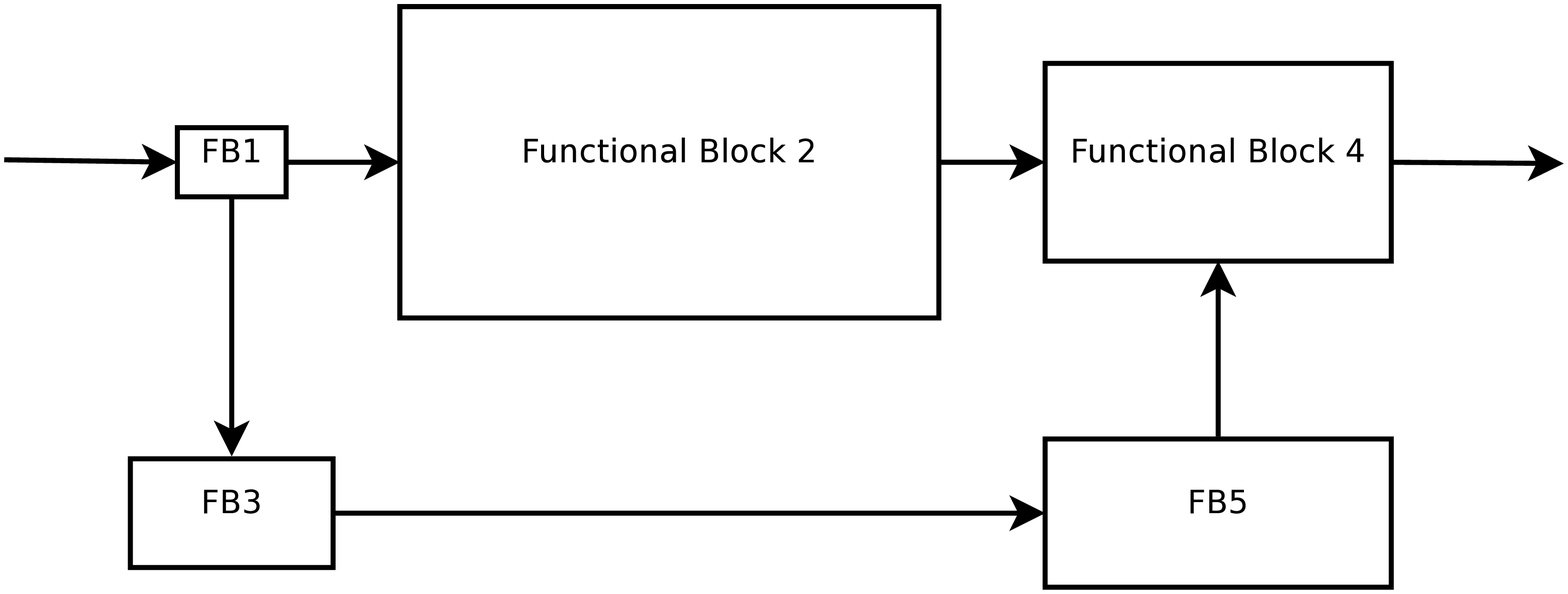,width=68.7mm} }
	\caption{Computational cost weighted functional block representation. Blocks 1 and 4 are \emph{peripheral}}
	\label{fig:chain2}
\end{figure}

We call $W_{n}$ the computational cost of the \emph{n-th} functional block and use a graphical representation of the SDR in which the size of the functional block is directly proportional to such cost, see figure \ref{fig:chain2}. It is possible to express $W_{n}$ either as the number of operations per second (Op/s) required by the block to work in real-time or as the CPU time it requires to process a given amount of data while being unthrottled (i.e. absorbing 100\% of available CPU resources while running). Obviously such representation choice has to be kept consistent throughout the MA design process.

Indeed we suggest the first method as the more practical when designing MA from paper (i.e. in absence of an optimized, computation-only reference implementation of the system that is being accelerated). Second method is instead best suited when such an implementation is available and CPU times absorbed by the single blocks can be measured. 

The symbol $Wm_{m}$ represents the total computational cost of memory management for table $t_{m}(...)$, that is the cost of memory address calculation and the equivalent computational cost (consumption of CPU time or of a given number of operations) yielded by memory access latencies, if any is present and significant.

$W$ is the total computational power in Op/s available within the computing platform.

$M$ is the total size of available memory within the computing platform, $M_{m}$ is the total memory footprint of table $t_{m}(...)$, $S_{m}$ is the data size of items stored in  $t_{m}(...)$, all such quantities are expressed in Bytes.

We call $\eta$ the acceleration efficiency for block $f(...)$: 

\begin{equation}\label{eq:eta}
\eta = \frac{\sum_{n=0}^{N_{sb}-1}W_{n}-\sum_{m=0}^{N_{t}-1}Wm_{m}}{\sum_{m=0}^{N_{t}-1}M_{m}}
\end{equation}

where $N_{sb}$ is the number of the obtained sub-blocks $f_{n}$ that will be implemented in memory through the use of suitable tables and $N_{t}$ is the number of tables used to produce such an implementation.

Acceleration efficiency [$\eta$] is therefore defined as the ratio between the computational effort being saved by means of the performed in-memory implementations, reduced by the amount of computational work needed for table management, and the total memory footprint being required.

It is also possible to define acceleration efficiency on a per-table basis, we obtain:

\begin{equation}\label{eq:etaperblock}
\eta_{m} = \frac{\sum_{n=0}^{N_{sb}-1}W_{n}-Wm_{m}}{M_{m}}
\end{equation}

where $N_{sb}$ indicates in this case the number of computational sub-blocks which are substituted by table $t_{m}(...)$. A negative value for $\eta$ (or of course $\eta_{m}$) has to be avoided as it indicates the chosen MA design will reduce system performance. 

\subsection{Acceleration Design}
\label{sec:AD}
Aim of Memory Acceleration is to assist the general purpose computing system in processing the informative signal (both Rx and Tx sides) through the usage of memory resources, thus increasing the overall throughput by a factor that we will call the \emph{acceleration factor}, in symbols: $a$.

Such result is obtained by convenient substitutions of functional blocks being conventionally implemented by using pure calculus $f_{n}(...)$ with the pre-computed tables $t_{m}(...)$ presented in paragraph \ref{subsec:repr} that is performed after one or more steps of \emph{algorithm segmentation} and according to the Recursive Table Aggregation Rule which will be described in the following. 

Therefore MA design reaches optimality when all memory resources available are used and the maximum possible value for $a$ is reached. That is, when all memory is used and is used well.

The \emph{input space cardinality} $[C_{i_{n}}]$ of each $f_{n}(...)$ is assumed to be \emph{independent} from its \emph{computational cost} $[W_{n}]$ or, in the worst case, weakly correlated. Think for example of three typical signal processing blocks being present in any modern radio system: a block-based Forward Error Correction (FEC) decoder, its associated interleaver and a scrambler preforming energy dispersal on a set of data sized precisely the same as the FEC block. Such three blocks share the very same $C_{i}$ but yield enormously different computational costs (i.e. at least one order of magnitude) with the FEC decoder being dramatically heavier than the other two.

Therefore, as long as the amount of memory resources $[M]$ is finite, it is necessary that those resources are used to accelerate the computationally-heaviest blocks of the chain. Performing memory-acceleration of a low $W_{n}$ block immediately would obliterate optimality as long as the memory used to replace computation yielded by such block could be better used in order to replace an heavier block.

Process reaches optimality when memory space is exhausted and the maximum possible number of operations (or the maximum possible amount of CPU time) has been replaced by memory look-ups providing the same output.

\subsection{Table aggregation and cache friendliness}
\label{sec:TAeCF}
\begin{figure}[h]
	\centerline{\psfig{figure=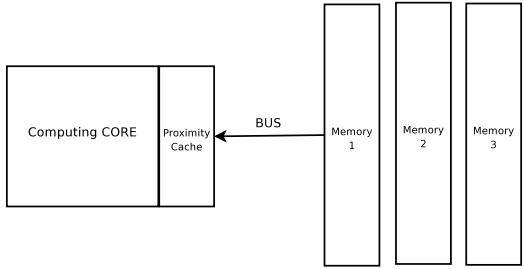,width=68.7mm} }
	\caption{High level memory structure on a general purpose computing system. Extended memory is typically made up of several blocks of RAM}
	\label{fig:memstruct}
\end{figure}

Considering that each implemented table yields, at least, one look-up act in order to perform the computation it is supposed to, we conclude that minimizing $Wm$ (i.e. the total memory management cost in our implementation) necessarily requires to aggregate as many functional blocks $f_{n}(...)$ as possible into a single table $t_{m}(...)$.
Also, we must consider that most general purpose computing systems have a hierarchical memory structure with smaller and faster caches in the proximity of the computing core and a bigger, slower extended memory in a more peripheral spot of the system (see figure \ref{fig:memstruct}). Therefore, it has a dramatic effect on system performance to store contiguously in memory information which will be used contiguously in time. Thus, as long as the functions implemented by each block are applied serially along the chain, it is a good thing that memory implementation of contiguous functional blocks happens within the same table $t_{m}(...)$. This actually makes it easy to structure the table in order to make it cache friendly, or either simply yields cache friendliness itself if cardinality of input spaces is small enough.

We thus propose a Recursive Table Aggregation Rule (RTAR) which is conceived in order to provide the aggregation of as many of the functional blocks as possible into the same table as well as in order to call for algorithm segmentation --which is a demanding task in terms of design effort-- only upon functional blocks where it is really needed and useful. 

\subsection{Recursive Table Aggregation Rule}
\label{subsec:RTAR}
A Recursive Table Aggregation Rule which ensures that MA design respects criteria discussed in \ref{sec:AD} and in \ref{sec:TAeCF}, provided that recursion is applied up to the exhaustion of memory resources, is presented in the following. RTAR is a recursive design algorithm for memory acceleration of a given SDR system.

Broadly speaking, it is possible to state that aim of algorithm segmentation is to break a functional block down into the highest possible number of component sub-blocks yielding as small input spaces as possible. Aim of RTAR is instead to re-aggregate as many of the \emph{heaviest} functional blocks (and sub-blocks), in which the system has been decomposed, within the smallest possible number of tables $t_{m}(...)$ as allowed by available memory resources.

Memory-contiguity of time-contiguous information as discussed in \ref{sec:TAeCF} is provided by RTAR as well as management of algorithm segmentation, which is invoked by RTAR only on the blocks where it is most effective.

We call \emph{table boundary} (TB) the line containing the subsystem we try to memory-accelerate during each step of the recursive design process.
Interfaces from and towards the remainder of the system are given by the system chain connection arrows that do cross the TB.
A block is called \emph{peripheral} if all of its input arrows and/or all of its output arrows do cross the \emph{table boundary}.

\begin{figure}[h]
	\centerline{\psfig{figure=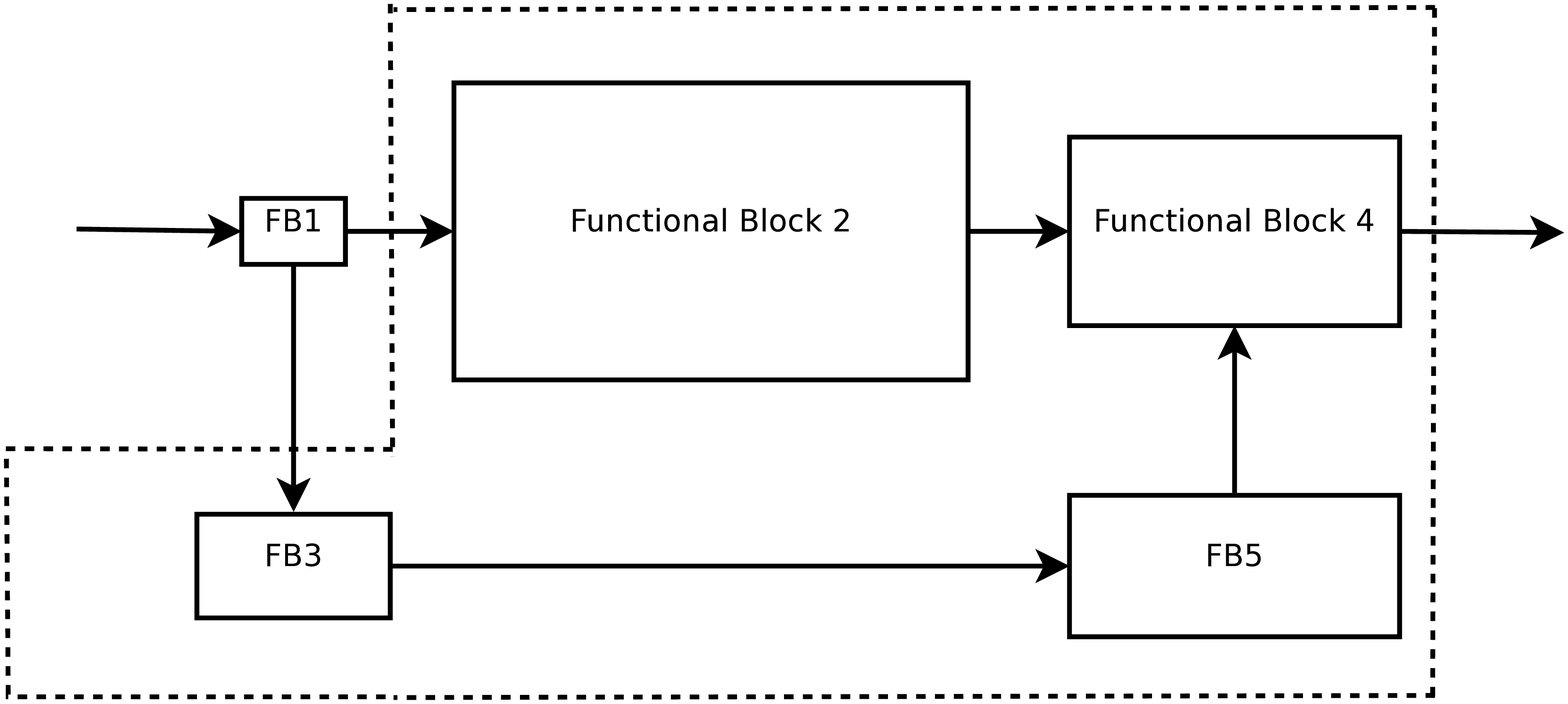,width=68.7mm} }
	\caption{Table boundary after one step, released block is peripheral}
	\label{fig:tb1}
\end{figure}
\begin{figure}[h]
	\centerline{\psfig{figure=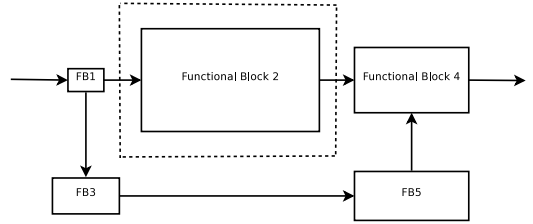,width=68.7mm} }
	\caption{Atomicity limit reached}
	\label{fig:atomreached}
\end{figure}
\begin{figure}[h]
	\centerline{\psfig{figure=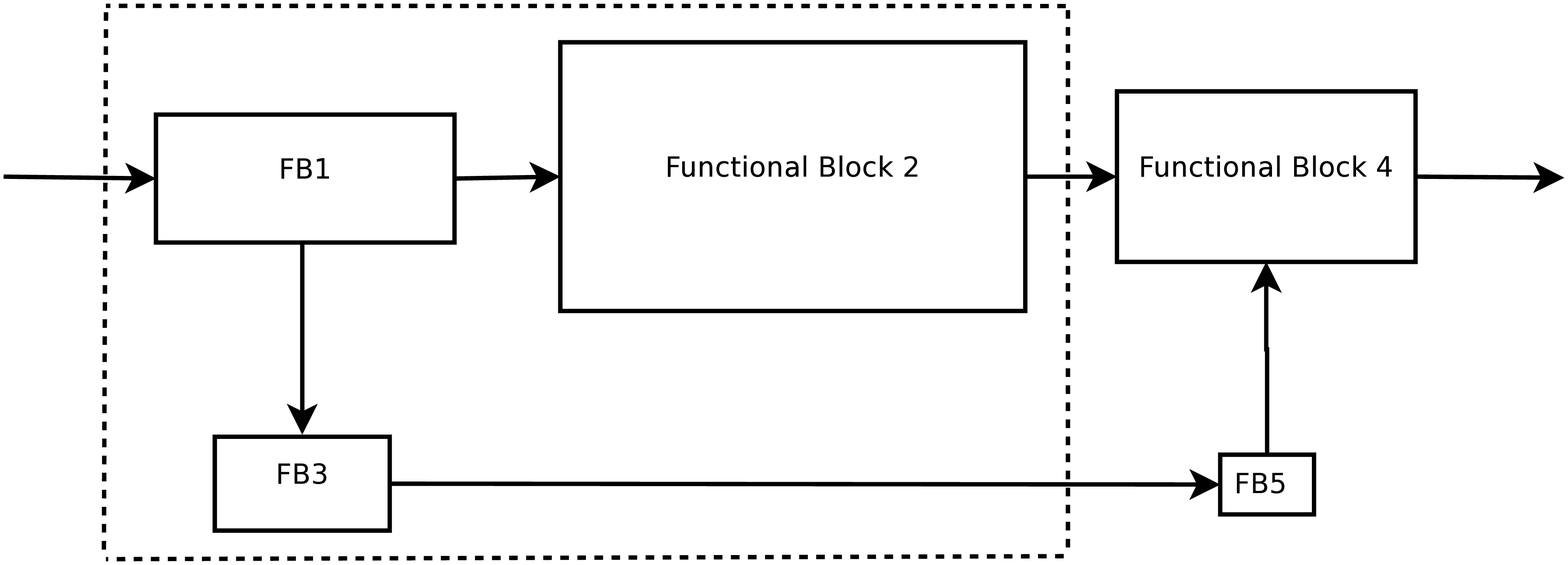,width=68.7mm} }
	\caption{Example of released block (FB5) being non-peripheral. In this case cascaded blocks (FB4) are released as well. At next step of the iterative algorithm, formerly released blocks will be enclosed in the new table boundary}
	\label{fig:tb2}
\end{figure}
\begin{enumerate}
\item{Define whole radio as sub-system to be memory-accelerated. This is equivalent to enclosing the entire radio within the table boundary. Calculate $SC_{i}$. If obtained table fits in memory, then go to step 3), if not, perform \emph{0-step} of algorithm segmentation. The table boundary now contains the system broken down to the block level.}
\item{Identify the computationally-lightest block contained within the table boundary and release it by moving it outside the table boundary (figure \ref{fig:tb1}). If the released block is not peripheral, then release all blocks depending on its output, see figure \ref{fig:tb2}. Calculate $SC_{i}$ again, if table fits then go to step 3), otherwise iterate step 2) up to when either the table fits or the $f_{n}(...)$ atomicity limit is reached (figure \ref{fig:atomreached}). If the latter becomes true, perform a further algorithm segmentation step over $f_{n}(...)$ and restart iterating step 2). Note that, in the case atomicity limit is reached, the block which will undergo algorithm segmentation is the heaviest block of the whole radio, then the sub-blocks obtained from segmentation collectively yield the majority of the computational cost of the SDR, subsequent iterations will thus be performed leaving the table boundary around such sub-blocks without re-initializing. Whenever one of the sub blocks obtained from algorithm segmentation is released, check whether the table boundary encloses a computational cost $W_{TB}$ which is greater than the cost of any functional block outside the TB. In case this condition becomes false, re-initialize the TB to enclose the entire system and iterate step 2). }
\item{Implement the subsystem being enclosed in the current table boundary by substituting its computation-only functional blocks with a suitable table $t(...)$. Table $t(...)$ will be an input/output map completely equivalent to the replaced subsystem. If there are still blocks not implemented in memory and memory resources are not exhausted, initialize table boundary for next iteration by enclosing all the remainder of the radio system, then go to step 2).}
\end{enumerate}
It must be noted that the proposed RTAR algorithm is sub-optimal. Indeed, aiming for optimality would require an exhaustive approach. Algorithm segmentation should then be performed on all blocks, regardless of their computational cost, then, on the completely algorithm-segmented version of the system to be memory accelerated, all possible aggregates of segments (i.e. the obtained sub-blocks) should be considered and characterized by means of saved computational cost ($W_{m}-W_{m_{m}}$) and memory footprint ($M_{m}$). After discarding all aggregates which are unable to fit in memory, the set of aggregates which maximizes $a$, under the system's memory resource constraints, should be implemented as memory tables.

Though suboptimal, we believe the proposed RTAR algorithm reaches a good trade-off point between performance and design complexity. In fact, it was able to provide substantial (roughly one order of magnitude) acceleration factors on the two very diverse algorithms presented in following sections, while yielding acceptable MA design effort.

\begin{figure}[h]
	\centerline{\psfig{figure=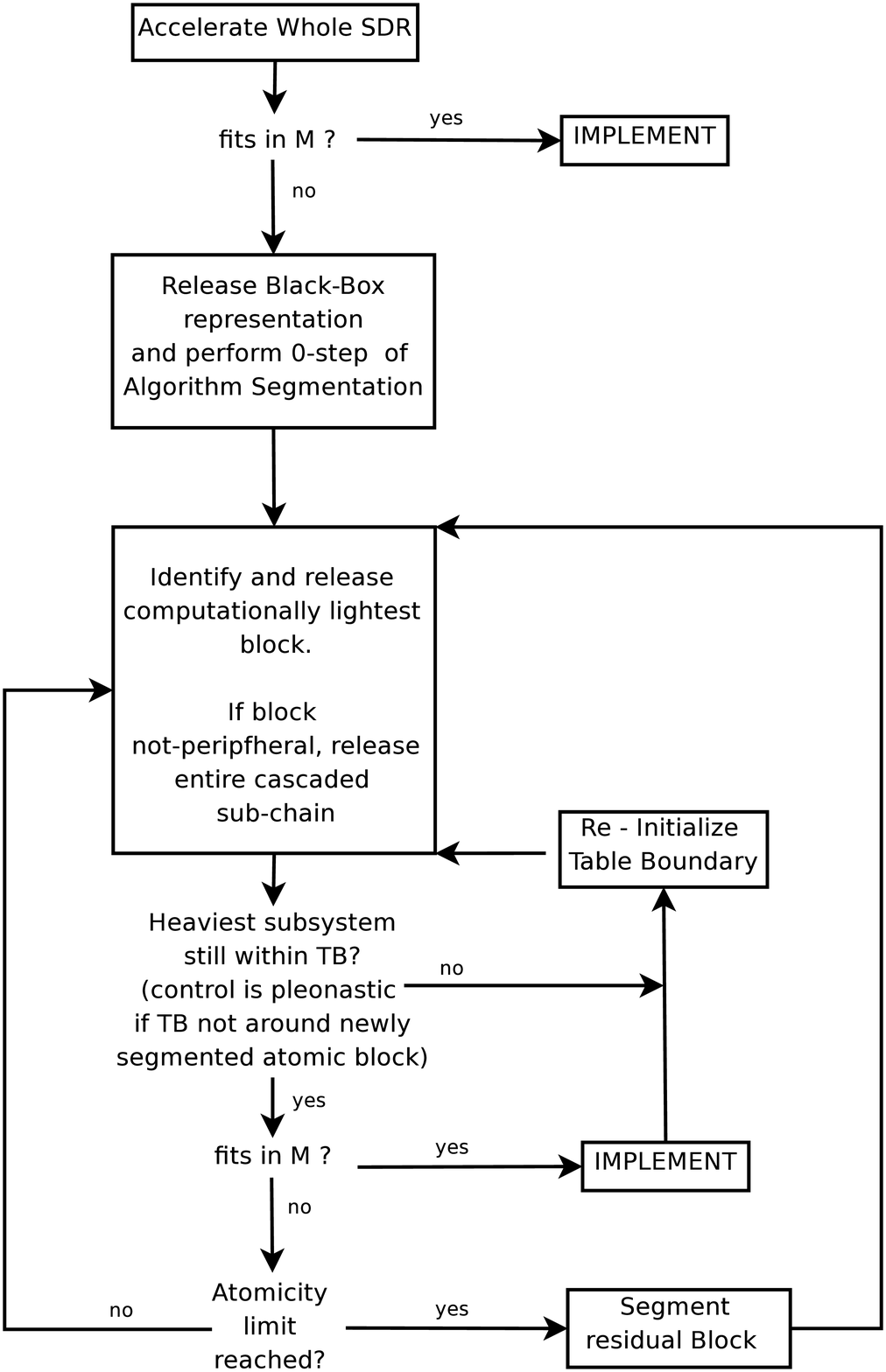,width=68.7mm} }
	\caption{Schematic representation of MA Recursive Table Aggregation Rule. Exit condition on memory exhaustion is not graphically represented for readability purposes}
	\label{fig:recalg}
\end{figure}

\subsection{More on Algorithm Segmentation}
\label{subsec:MAS}
As previously stated we call \emph{Algortim Segmentation} the process of breaking a single functional block $f(...)$ up into its constituent functional sub-blocks or \emph{segments}. 

Algorithm Segmentation simply identifies segments within the given algorithm $f(...)$ while doing no re-design of the segmented algorithm. Thus, computational cost of segmented system $f(...)$ is conserved.

A \emph{segment} is any sub-system of $f(...)$ for which a MIDS can be identified (over one or more input lanes). Output yielded by the processing of such MIDS is input to one or some of the subsequent segments which concur to build up $f(...)$ as a whole.

The reason for which algorithm segmentation is such a crucial tool for MA is that there is \emph{positive correlation} between the \emph{number of different sub-functions} that a functional block performs and $l$. This happens because, within a functional block that includes many sub-blocks, such sub blocks do operate with different \emph{granularities} (i.e. with different MIDS) and therefore the MIDS of the entire block typically grows, in order to accommodate all of the underlying, to their \emph{least common multiple.}

As a consequence, being $C_{i}$ an extremely non linear function of $l$ (indeed, it is exponential in $l$) as shown in \ref{eq:Ci}, it is highly convenient to keep $l$ as small as possible by means of algorithm segmentation in order to obtain tables $t_{m}(...)$ which can fit into the available memory resources.

In practice it turns out that functional blocks that do perform several \emph{different} sub-functions will have a large $l$ and will require segmentation in order to be (even partially) implemented in memory.
Please note that we are not stating the presence of any kind of correlation between input space cardinality $C_{i}$ and computational cost of a block.  

As radio signal processing algorithms do differ really much from one another, they do offer very different opportunities for algorithm segmentation. It is therefore difficult to give optimality bounds for algorithm segmentation into an MA context, still we can say that the best algorithm segmentation is the one providing the finest possible \emph{granularity of input spaces} of obtained sub blocks. This is true because the smaller the granularity is, the closer the RTAR will manage to bring the total memory occupancy of the MA-ed SDR, $\sum_{m=0}^{N_{t}-1}M_{m}$, to the memory capacity of the system $M$. 

Thus, for what stated above about the positive correlation between number of sub-functions implemented by a block and its MIDS size, it turns out that the more sub-blocks $N_{sb}$ algorithm segmentation obtains from the given block, the better algorithm segmentation was performed.

Finally it is possible to state that the joint action of algorithm segmentation and RTAR is to \emph{decompose the given SDR system down to the finest possible level of computational granularity, in order to allow for a re-implementation which is capable of using as much as possible of the available memory resources to perform the heaviest part of the computation that the SDR requires. All this, just with the smallest possible computational cost of memory management.} 

This is the gist of the MA concept.

\subsection{Some more MA Analytics}
Once the acceleration process has been completed (i.e. the RTAR algorithm terminates for memory exhaustion or on having memory-accelerated the entire system) it is possible to calculate the obtained acceleration factor $a$ as
\begin{equation}\label{eq:a}
a = \frac{W_{r}+\sum_{n=0}^{N_{sb}-1}W_{n}}{W_{r}+\sum_{m=0}^{N_{t}-1}Wm_{m}}
\end{equation}

where $N_{sb}$ is the number of functional sub-blocks $f_{n}(...)$ that have been implemented in memory through the use of suitable tables and $N_{t}$ is the number of tables that have been used to produce such an implementation. $W_{r}$ accounts for the computational cost of remaining blocks which where not implemented in memory.

It is indeed very difficult to give an upper bound for $a$, that is determine the optimal acceleration which can be obtained through MA, as long as such acceleration depends on optimality of algorithm segmentation, which in turn heavily depends on the algorithm (the functional block) which is to be segmented and memory-accelerated. Different algorithms offer different opportunities for segmentation and therefore $a_{max}$ is highly dispersed among implementations. Still, an estimate can be given, which though is fully dependent on the number of tables $N_{t}$ that the RTAR obtains from the segments $N_{sb}$ the given system $f(...)$ (either a functional block or the entire radio) has been broken up into by \emph{algorithm segmentation}. It is assumed that the complete system $f(...)$ finally fits into the available memory.  We have:
\begin{equation}\label{eq:U}
a_{max} = \frac{\sum_{n=0}^{N_{sb}-1}W_{n}}{\sum_{m=0}^{N_{t}-1}L_{m}+(i_{m}-1)(x + \sigma)}
\end{equation}
where $L_{n}$ is the access latency for each table (depending on table dimension and chosen implementation platform), $i_{m}($ is the number of inputs to each table, $x$ is the computational cost for one multiplication by a constant and $\sigma$ is the computational cost of one sum with a variable.
All such quantities, including $L$, can be both expressed in terms of number of operations and required CPU time.

The numerator is the total computational cost of the original system $f(...)$ expressed as the sum of computational costs of all sub-blocks it was broken down into. The denominator general term $L_{m}+(i_{m}-1)(x + \sigma)$ is indeed our estimate for $Wm_{m}$.

As already stated, aim of MA is to create software radios that exploit all the resources available on a general purpose computing system, that is not only calculus but also memory. Thus, based on the acceleration efficiency, it is possible to derive an overall merit parameter which describes quantitatively how a memory-accelerated SDR implementation ranks with respect to another one.
Given a radio whose black-box aggregate representation we call $f(...)$, we define $I$ as the overall merit parameter of the memory-accelerated implementation of $f(...)$.

\begin{equation}\label{eq:I}
I = \frac{\sum_{m=0}^{N_{t}-1}M_{m}}{M}\frac{W_{impl}}{W}\frac{\sum_{n=0}^{N_{sb}-1}W_{n}-\sum_{m=0}^{N_{t}-1}Wm_{m}}{\sum_{m=0}^{N_{t}-1}M_{m}}
\end{equation}

Where $W_{impl}$ represents the computational work required by the implemented radio when running throttled, that is on line and processing exactly the amount of samples per second which is required by the chosen radiofrequency communication standard. Actually the first two rational factors do account for use of all available memory and computational resources respectively. 

A good way of evaluating $I$ for the various MA-SDRs one might wish to compare is to run the radio unthrottled (i.e. at the full speed allowed by the hardware it is running upon), this yields:

\begin{equation}\label{eq:unthrottle}
\frac{W_{impl}}{W} = 1
\end{equation}

Assuming this and simplifying the total memory occupancy of the tables $\sum_{m=0}^{N_{t}-1}M_{m}$ in \ref{eq:I} we obtain

\begin{equation}\label{eq:I2}
I = \frac{1}{M}\left(\sum_{n=0}^{N_{sb}-1}W_{n}-\sum_{m=0}^{N_{t}-1}Wm_{m}\right)
\end{equation}
 
\subsection{MA Compatibility}
All performance results presented in section \ref{RIMEM} have been obtained by applying MA alone, i.e. by making use of no other performance enhancement technique such as low level (Assembler) programming or parallelization. This was done in order to explore the contribution to performance enhancement that MA can provide by itself.

Still, it is important to note how MA is fully compatible with such performance-computing typical implementation techniques and to consider how such techniques could provide further acceleration when applied within an MA context.

\begin{figure}[h]
	\centerline{\psfig{figure=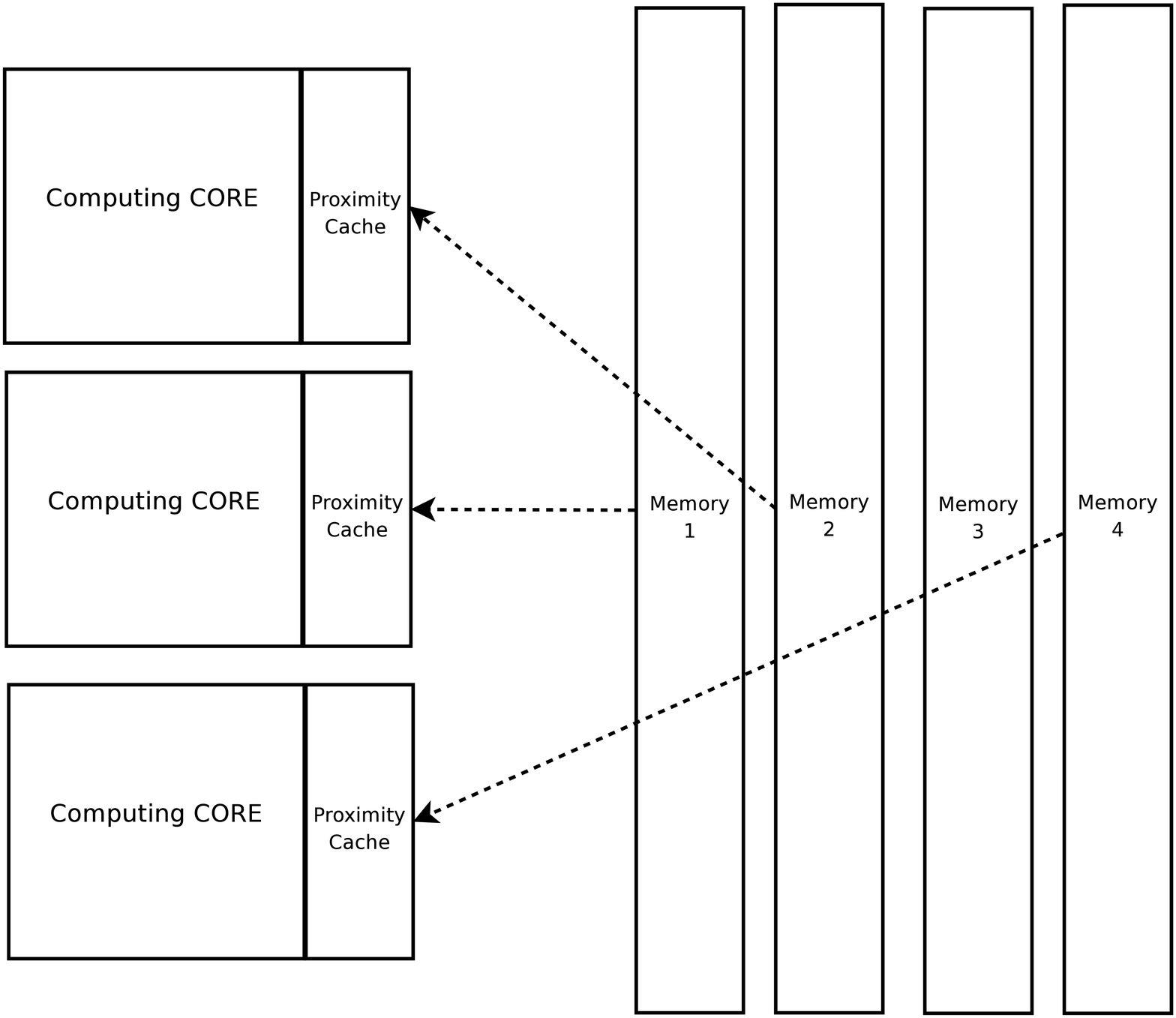,width=68.7mm} }
	\caption{Multiple processing cores, their dedicated caches and table loading from RAM to the core cache}
	\label{fig:multicorecache}
\end{figure}

Compatibility with low level programming is trivial and does not deserve a discussion. For parallel implementation instead, a possible objection is that concurrent access to a certain memory area from multiple computing cores can result in the need for collision control and, therefore, performance bottleneck. Such problem can be easily avoided by simply making use of cache friendliness. Multicore computing systems do often have cache memories which are dedicated to each single core as depicted in figure \ref{fig:multicorecache}. Such memories are independently accessed by each of the cores removing the possibility of collisions.

Collisions could instead happen when loading the required memory table (or table portion) from the external Random Access Memory (RAM) into the core-dedicated caches as shown in figure \ref{fig:multicorecache}. Appropriate use of a cache-friendly table structure will make these fetches extremely rare, while the frequent look-ups required by our memory-based computation scheme happen only within the core-dedicated cache. Therefore, should access contention happen at the RAM-level, it would be rare enough not to threaten the performance level of the system.   

\section{RIMEMBRI: \\MA tested on an everyday road} 
\label{RIMEM}
\subsection{MA design for Viterbi decoder} 
As a real-world \emph{proof of concept} of MA we provide the results we obtained by applying the MA technique to our fully software implementation of a an ETSI DVB-T \cite{ETSI_DVBT} receiver called R-DVB \cite{ROSE_THESIS}.

After implementing R-DVB the usual, computation-only, way and performing several cycles of code optimization, our realtime distance ratio was about 10 (i.e. R-DVB requiring 10 times the realtime in order to process a given amount of samples), R-DVB including the entire ETSI DVB-T reception chain from reference signals removal up to demodulated Transport Stream (TS) delivery. Test and reference platform for all further considerations is an Intel Q9400 CPU clocked at 2.66 GHz. All implemented software makes \emph{no use of parallelization} (all code is single threaded), therefore \emph{only one of the four cores} available on our test CPU is used by any of the implementations presented here.

Obviously, aiming to implement an entire ETSI DVB-T receiver as a single memory table (i.e. by applying no algorithm segmentation nor any RTAR cycles), makes no sense at all, still, we assumed such ideal wish as the formal starting point of RTAR. Once released the single block representation of the system, we profiled the computational cost of the entire demodulation chain on a functional-block basis, thus performing what we called the \emph{0-step} of algorithm segmentation.
It was immediately clear that the K=7 Viterbi decoder \cite{VITERBI} included in the demodulation chain \cite{ETSI_DVBT} was by far the heaviest block of the system. Actually, Viterbi alone took about $8$ times the realtime, while the remainder of the chain took only $0.5$ times the realtime to be executed. This meant, according to RTAR, that Viterbi decoder was meant to be the first block to undergo memory acceleration. Such activity produced a novel implementation of the Viterbi decoder algorithm, relying almost entirely upon memory resources, which we then humorously named \emph{RIMEMBRI} in order to stress its peculiar, memory-based nature.

Such memory implementation of the by far heaviest block of the receiver chain was what enabled us to obtain a completely realtime, fully software ETSI DVB-T receiver on a low budget general purpose CPU as described in \cite{SR-DVB_WSR10}. Such a result would not have been possible without MA. \emph{Rimembri} implementation is described in the following. 

\begin{figure}[h]
	\centerline{\psfig{figure=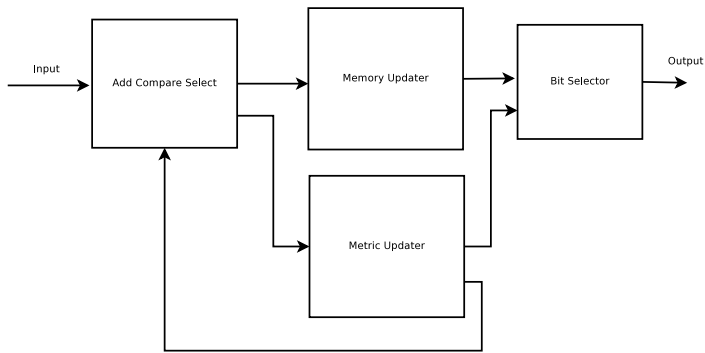,width=92.7mm} }
	\caption{Classical functional block decomposition of Viterbi Decoder. All displayed blocks are implemented through pure computation}
	\label{fig:vit-classic-decomp}
\end{figure}

Actually, as stated above, applying RTAR recursively as described in \ref{subsec:RTAR}, we got to the point where we were requested to segment our ETSI DVB-T, K=7 Viterbi decoder. A classical functional block decomposition of the Viterbi algorithm is shown in figure \ref{fig:vit-classic-decomp}, where all functions are implemented through pure computation.
The blocks of such decomposition are the usual Viterbi decoder Add Compare Select (ACS) function, a block updating the memories of decoded bits (path bitsets) for each Viterbi decoder state, a block updating the metrics (weights) for each of the decoder's state and a block selecting the likeliest (smallest metric) state after a decoding-depth long observation.

\begin{figure*}[h]
	\centerline{\psfig{figure=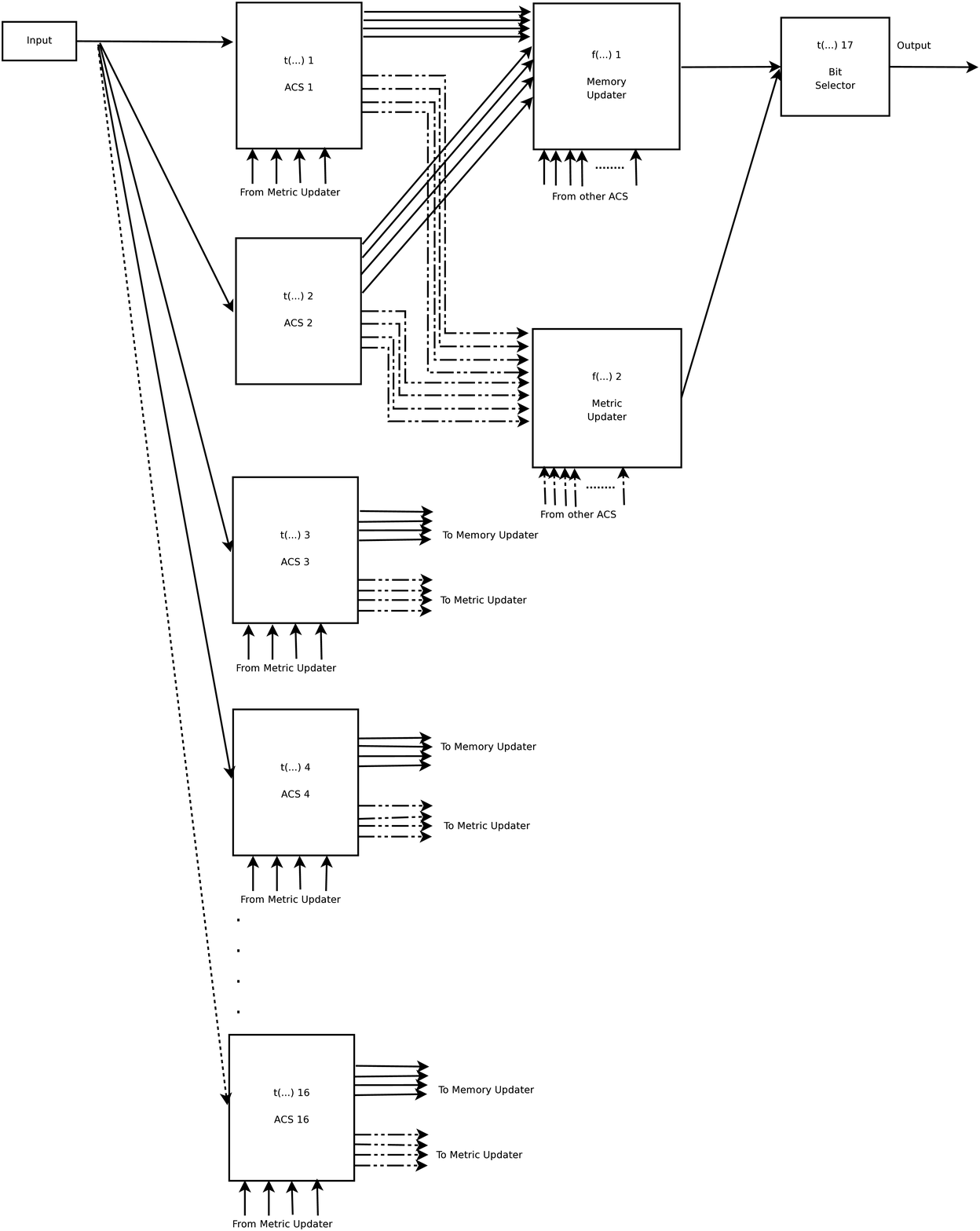,width=150.7mm} }
	\caption{Memory-accelerated Viterbi decoder as returned by RTAR. f(...) indicates a purely computational implementation, t(...) a memory table}
	\label{fig:vit-RTAR}
\end{figure*}

Obviously, after such basic segmentation, input space cardinalities $C_{i_{n}}$ were still far too large to fit into available memory. We iterated RTAR until we obtained the segmented implementation of figure \label{fig:vit-AS} and convenient table boundaries. Within such implementation, a much finer segmentation of the Viterbi algorithm is visible. Previous functional blocks have been broken up into several constituent sub-blocks, namely:
\begin{itemize}
\item{Calculation of current input bitset's distance from each transition label}
\item{Sum of branch metrics}
\item{Branch metric comparison and selection}
\item{Sum and updating of path metrics}
\item{Updating of path bitsets}
\end{itemize} 
Upon each RTAR iteration completion, a table boundary was produced while algorithm segmentation was performed when required. By implementing each table boundary (i.e. the functional blocks aggregated within) as a memory table, we obtained the implementation shown in figure \ref{fig:vit-RTAR}. In such implementation, Add, Compare and Select functions for as many as 4 states are performed through a single memory look up. Another memory table implements the selection of the likeliest state, while update of decoded bit memories and of current metrics for each state is still being performed through pure computation. 

As memory resources of our test system are being just marginally exploited (50 MiB out of 4 GiB) and some functions are still implemented the usual, computation-only way, RTAR could be further iterated in order to bring to memory also such functional blocks, providing further acceleration to the overall system.

The relevant implication of the described implementation strategy is that the computational cost required to perform the single memory look-up is by far smaller than what direct computation of the required result would yield. For this reason we claim that the described approach \emph{aggregates many elementary computation acts within a certain number of CPU cycles} and thus reduces the power efficiency gap between SW and HW implementations, without losing anything in reconfigurability and flexibility of the system. 

\begin{figure*}[h]
	\centerline{\psfig{figure=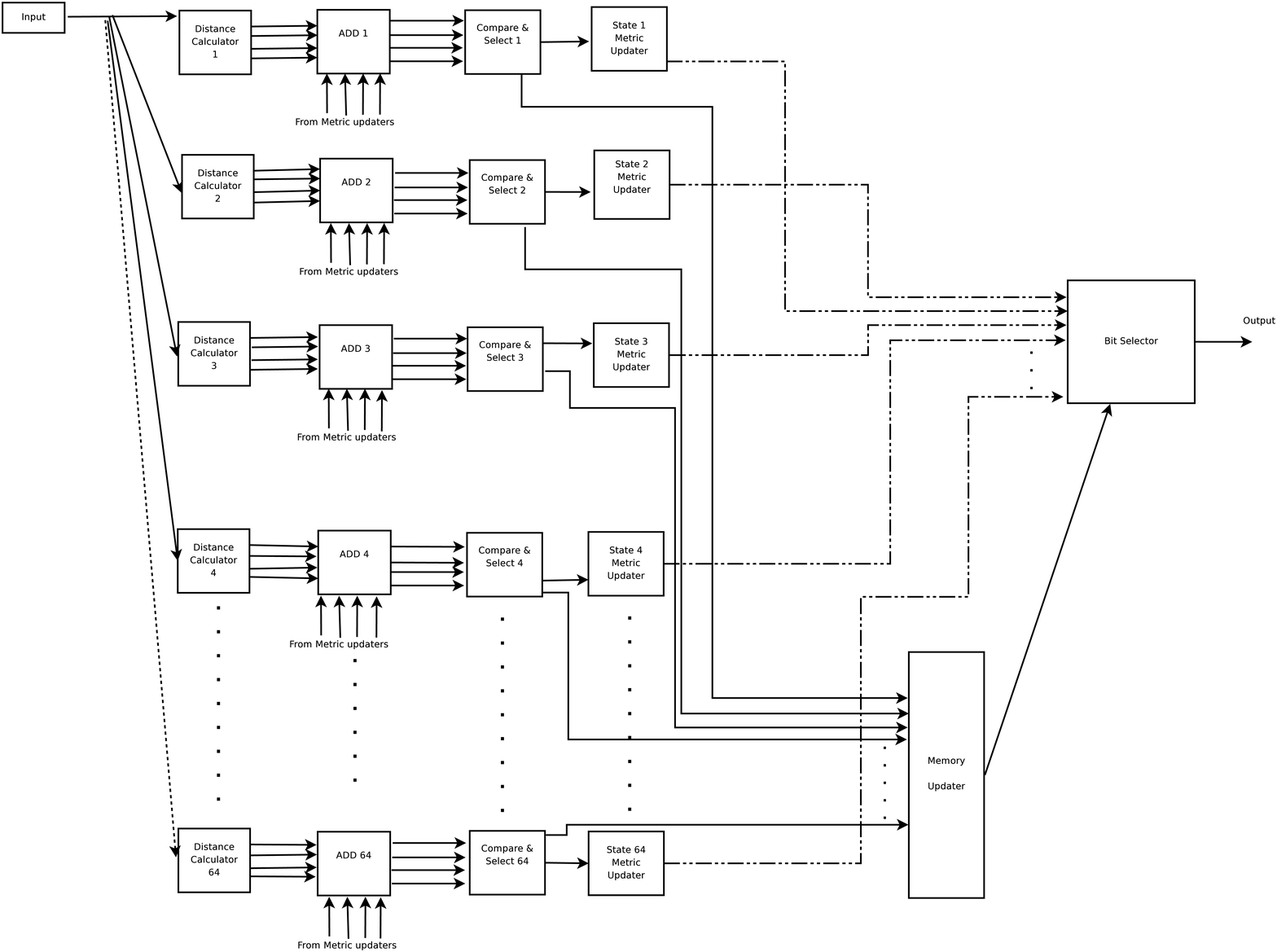,width=150.7mm} }
	\caption{Computation-only implementation of our Viterbi decoder, right after undergoing algorithm segmentation. f(...) indicates a purely computational implementation}
	\label{fig:vit-AS}
\end{figure*} 

\subsection{Performance results} 
On the reference CPU described in section \ref{RIMEM}, the computation-only implementation of our Viterbi decoder takes about $7.71$ times the realtime. After undergoing memory acceleration, the same implementation takes $0.74$ times the realtime. Acceleration factor $a$ equals $10.4$.
Total memory occupancy of the memory-accelerated implementation is $50.0$ MiB. Presented performance test results were obtained by compiling sourcecode with \emph{g++} compiler, version $4.3.2-7$.

\section{Other memory-accelerated algorithms} 
\begin{figure}[h]
	\centerline{\psfig{figure=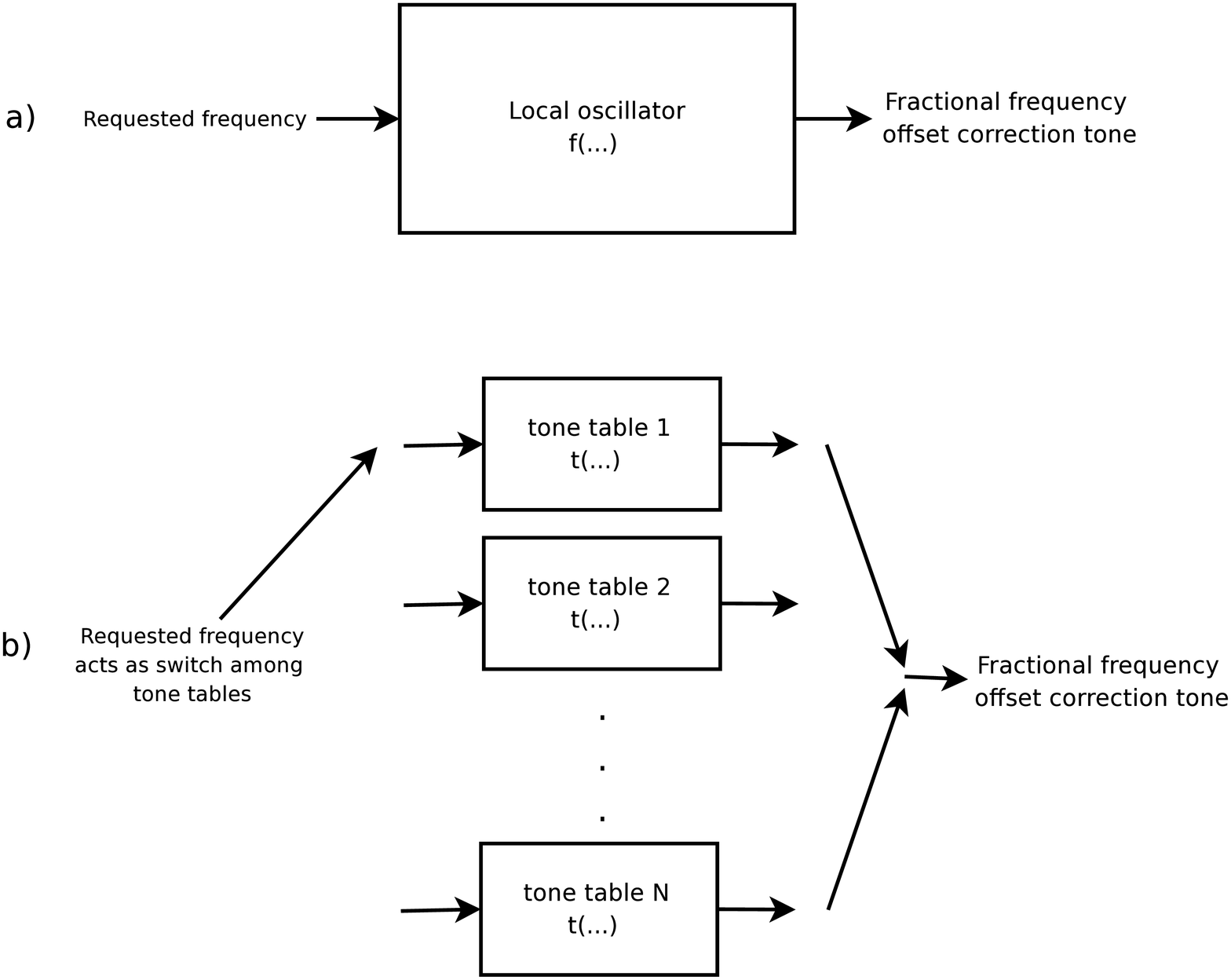,width=76mm} }
	\caption{A very basic MA application: segmentation of a computationally-implemented local oscillator a), into a finite set of memory-tabled complex oscillations b)}
	\label{fig:MA_LO}
\end{figure}

Some results obtained while applying MA to other algorithms are worth to be mentioned as well. Within the synchronization section of the same ETSI DVB-T receiver (please see \cite{DIDIO_THESIS}) a carrier frequency fractional (i.e. expressed as a fraction of OFDM subcarrier spacing) offset corrector which used to be implemented by means of pure calculus was accelerated by a factor $46.3$ after undergoing MA. In short, within this very basic MA application, a single, computationally-implemented, oscillator generating any complex tone that could be required to compensate the estimated fractional offset is \emph{algorithm-segmented} down to a set of oscillators capable of generating only a single frequency. Each of such sub-blocks (sub-oscillators) is implemented by means of a memory table, therefore obtaining full memory implementation of the fractional frequency offset corrector block as shown in figure \ref{fig:MA_LO}.

As discussed above, functional blocks inherently working on large amounts of data (i.e. featuring big MIDS) require well-designed segmentation in order to be conveniently accelerated. When such blocks do perform very basic operations on such broad data sets, they are difficult to segment. At the time of writing this article, work is underway in order to obtain the largest possible memory-acceleration of quite a computationally heavy algorithm presenting such challenges, namely the OFDM time and frequency offset estimator described by van de Beek, Sandell and Borjesson in \cite{BORJ}. Memory-accelerated implementations obtained so far, provided acceleration factors as big as $11.996$ which are expected to grow by means of further development and will be described in future MA-related works. Reference HW/SW platform for this implementation is the same as described above.

We believe that, by having obtained acceleration factors of about one order of magnitude in terms of computational efficiency by applying MA to two highly diverse signal processing algorithms (namely an hard-valued Viterbi decoder and a soft-valued, correlator-based OFDM synchronizer), we have shown the generality of MA approach, that is its applicability to an extremely wide variety of radio signal processing algorithms. 

For such a reason, authors do propose MA as an implementation technique for radio signal processing within SDR systems which can provide a substantial boost to their performances and therefore move SDR technologies much closer to market segments they are currently excluded from, because of their poor energy efficiency.

\section{MB-SDR: \\Memory-accelerating  the entire radio}  
Based on the performance boost obtained by applying MA approach to our radios, curiosity naturally arises about evaluating the possibility to take the MA concept to its absolute limit: \emph{memory-implementation of the entire radio system}.
 
At this point it might be useful to discuss \emph{what} actually distinguishes software implementations from hardware ones to the extent that average power efficiency gap between the two is as wide as two orders of magnitude in favour of HW systems (with worst cases reaching three orders of magnitude).
Key concept in order to understand this huge separation is \emph{specificity}. Software implementations do rely on general purpose processors, which means that such computing systems feature as-small-as-possible operation granularity. I.e. the simplest arithmetic operations (sums) are \emph{serially} performed a huge number of times and combined in order to produce the required processing. This yields that a given (rather complex) operation will require a very big number of clock cycles to be performed. On the other hand, HW implementations consist of task-specific circuits
made up of many synchronous logic components. Thus, in a HW implementation, a large number of elementary operations will be \emph{aggregated within a single clock cycle} (i.e. each time the system clock signal completes a cycle, many elementary operations take place throughout the entire system, in all of its sections), clock frequency then can be kept much lower than what is required by SW systems in order to perform, in real-time, the amount of computation required by a certain radio standard. This is actually where the power efficiency gap between the two implementation classes is generated.

In section \ref{sec:AD} we described MA technique as a way of assisting processing-yielded computation through the usage of memory resources. Indeed, given the above analysis of the difference between HW and SW systems, we could even look at the MA approach as a way of making up for the low operation granularity that characterizes software systems and impairs them with respect to their HW counterparts, when it comes to comparing power efficiencies. I.e. as HW systems do aggregate many operations within the same clock cycle by means of implementing several dedicated subsystems all triggered by each and every clock edge, \emph{MA approach aggregates many elementary operations by storing the result of their combination into a single table}. Some clock cycles will then be used in order to access the content of the table but still, if MA was designed correctly, the amount of \emph{equivalent} elementary operations \emph{performed per single clock cycle} will be greatly increased. Based on such considerations, the idea of extending memory implementation to the entire radio system starts getting attractive. 

Thus, we define a Memory Based SDR (MB-SDR) as \emph{a software defined radio where most (or all) of the computation is performed by means of suitable memory look-ups}. It might appear at a first glance that such a design choice conflicts with what stated in section \ref{subsec:repr}, where we claimed that efficient SDR implementations must use up all resources available (calculus and memory) in order to perform their computation. Actually, by implementing an MB-SDR, calculus resources are not left unused, they are used (exclusively or almost exclusively) to cover the computational work [$W_{m}$] yielded by memory management. This obviously loosens performance requirement over computing resources and therefore \emph{mandates downsizing of the computing core} in order to have it fully loaded and obtain the required increase in power efficiency.
As a result, role of the computing core of an MB-SDR is just to \emph{move the data around} and \emph{fetch the necessary information} from the right memory tables while actual computation truly happens only \emph{in memory}. Considering that, with present technology, an average computer CPU can take about 140 Watts of electrical power while 2 GiB, Double Data Rate 2 (DDR2) RAM modules typically require 4.4 to 5 Watts, this appears to be an interesting strategy to increase power efficiency of SDRs.

It is important to note how such a power efficiency gain has absolutely no impact upon the flexibility of the system. The flexibility and ease of reconfiguration which are peculiar to any SDR are fully conserved by the MA/MB-SDR approach as long as resources used for speeding up the computation are provided by memory and not by a different (less flexible) computation technology, as it happens instead whenever performance boost is gained by replacing SW implementations with specialized HW.

As long as we have defined above an MB-SDR as accommodating in memory \emph{all}, but even just \emph{most}, of the computation yielded by the radio communication system, hierarchical, computational-cost-driven memory-mapping of functional blocks remains a necessity. RTAR is thus kept as the instrument of such prioritization while algorithm segmentation still provides input space cardinality reduction as described for the general case in subsection \ref{subsec:MAS}. 

\section{Conclusions and perspectives}
Based on the here-described results obtained by applying MA to a considerably broad set of radio signal processing algorithms, we believe the MA technique presented here can provide substantial performance boost, and thus power efficiency, to existing SDR systems as well as to any system performing signal processing functions over general purpose computing architectures.
As long as different algorithms dramatically differ in the opportunities they offer for algorithm segmentation (and thus in the achievable acceleration factor) there is no algorithmic solution to perform AS, which therefore remains a peculiarly human design task. For such reason, we suggest this work opens an interesting lane for research into the signal processing field. Finding optimal segmentation of classical radio signal processing algorithms in order to allow for the most efficient memory acceleration is both a challenging effort and a promising research path. Actually, upon success, performance boosts of at least \emph{one order of magnitude} could be delivered to existing Software Defined Radios by making them energy-practical \emph{without reducing their peculiar flexibility/reconfigurability features}. Both theoretical and applied signal processing skills are needed in order to move along such research path. 

It must be also considered that acceleration factors presented within this work are obtained by using both computing architectures and compilers (GNU g++) that are totally unaware of the MA approach and therefore neglect memory access optimization in favour of pure computation. It is thus expected that applying MA on computational back-ends that take into account memory management and access optimization would result in even bigger acceleration factors.

Though developed for general purpose CPUs, Memory Acceleration can easily be applied to other computational architectures, for example multicore Digital Signal Processors (DSPs) or Multiprocessor System-on-Chip (MPSoC) implementations. To such systems it would provide substantial acceleration by offloading computational effort from processing cores towards onboard memory resources.

In conclusion, we might state that MA / MB-SDR approach promises to be a considerable step along the path to truly flexible radios that have very small loss in power efficiency with respect to equivalent hardware implementation.

\hfill March 31, 2010

\ifCLASSOPTIONcaptionsoff
  \newpage
\fi



\bibliographystyle{IEEEtran}

\begin{thebibliography}{}
\providecommand{\url}[1]{#1}
\csname url@samestyle\endcsname
\providecommand{\newblock}{\relax}
\providecommand{\bibinfo}[2]{#2}
\providecommand{\BIBentrySTDinterwordspacing}{\spaceskip=0pt\relax}
\providecommand{\BIBentryALTinterwordstretchfactor}{4}
\providecommand{\BIBentryALTinterwordspacing}{\spaceskip=\fontdimen2\font plus
\BIBentryALTinterwordstretchfactor\fontdimen3\font minus
  \fontdimen4\font\relax}
\providecommand{\BIBforeignlanguage}[2]{{%
\expandafter\ifx\csname l@#1\endcsname\relax
\typeout{** WARNING: IEEEtran.bst: No hyphenation pattern has been}%
\typeout{** loaded for the language `#1'. Using the pattern for}%
\typeout{** the default language instead.}%
\else
\language=\csname l@#1\endcsname
\fi
#2}}
\providecommand{\BIBdecl}{\relax}
\BIBdecl

\end{thebibliography}


\begin{thebibliography}{1}

\bibitem{STTO:1}
D.~Cociorva and G.~Baumgartner and C.G.~Lam and P.~Sadayappan and J.~Ramanujam and M.~Nooijen and D.E.~Bernholdt and R.~Harrison,
\emph{Space-Time Trade-Off Optimization for a Class of Electronic Structure Calculations}, New York, USA: ACM SIGPLAN Notices, Volume 37, pages 177-186, ACM, 2002
\bibitem{STTO:2}
C.D.~Walter, \emph{Space/Time Trade-Offs for Higher Radix Modular Multiplication Using Repeated Addition}, IEEE TRANSACTIONS ON COMPUTERS, VOL. 46, NO. 2, FEBRUARY 1997
\bibitem{STTO:3}
M.~Stamp, \emph{Once Upon a Time-Memory Tradeoff}, available online: http://en.wikipedia.org/wiki/Space-timetradeoff/\#External\_links

\bibitem{ETSI_DVBT}
  \emph{Digital Video Broadcasting (DVB); Framing Structure, channel coding and modulation for digital terrestrial television}, ETSI EN 300 744 V1.5.1, Sophia Antipolis, France. 
\bibitem{VITERBI}
A. J. Viterbi, \emph{Error Bounds for Convolutional Codes and an Asymptotically Optimum Decoding Algorithm}, IEEE Transactions on Information Theory , vol. IT-13, April, 1967, pp. 260-269
\bibitem{ROSE_THESIS}
 L. ~Rose and M. ~Luise and F. ~Giannetti and V. ~Pellegrini,\emph{R-DVB: Software Defined Radio implementation of DVB-T signal detection functions for digital terrestrial television}
\bibitem{SR-DVB_WSR10}
V.~Pellegrini, M.~Di Dio, L.~Rose, M.~Luise \emph{``A real time, fully software, ETSI DVB-T receiver based on the USRP,''} in proceedings WSR10, Karlsruhe, Germany, March 2010.
\bibitem{DIDIO_THESIS}
 M. ~Di Dio and M. ~Luise and F. ~Giannetti and V. ~Pellegrini, \emph{Signal synchronization and channel estimation/equalization functions for DVB-T software-defined receivers}
\bibitem{BORJ}
 J. J. van de Beek and M. Sandell and P. O. Borjesson, \emph{ML estimation of time and frequency offset in OFDM systems}
\end{thebibliography}
%

%

\begin{IEEEbiographynophoto}{Vincenzo Pellegrini}
Vincenzo Pellegrini received the B.E.
degree in telecommunications engineering from the University 
of Pisa, Pisa, Italy, in July 2006.
Since the beginning of his B.E. degree thesis, he is actively 
working with the ``DSP for Communications Lab'' (DSPCOLA) of 
the Dept. of Information Engineering, University of Pisa, 
under the supervision of Prof. Marco Luise.
From February 2008 to December 2008 he worked as the main software developer and hardware sub-system integrator of European project GRIDES at Wiser Srl, Livorno.
In January 2008 he obtained the first fully software radio implementation of an ETSI DVB-T modulator capable of running realtime over low cost general purpose CPUs. Results of such work were published along with a full system demonstration in march 2008 at WSR08 conference in Karlsruhe, Germany. 

In September 2008 he received his cum-laude M.E. degree from the same university.
 
Since January 2009 he is attending his PhD studies in radio transmission systems, still at Pisa University. 
\end{IEEEbiographynophoto}

\begin{IEEEbiographynophoto}{Luca Rose}
Luca Rose received his cum-laude M.E. degree from Pisa University in April 2009. He has been working within DSPCoLa up to September 2009 developing the memory-accelerated Viterbi decoder described within this work. He is now with Sup\'{e}lec, Paris. 
\end{IEEEbiographynophoto}

\begin{IEEEbiographynophoto}{Mario Di Dio}
                         received his cum-laude
                         B.E. degree in telecommunications en-
                         gineering from the University of Pisa,
                         Pisa, Italy, in July 2007 with a the-
                         sis on wireless cooperative communica-
                         tion. In September 2009 he received
                         his cum-laude M.E. degree from the
                         same university with a thesis developing
                         the synchronization and channel estima-
tion/equalization functions for a SDR fully-software DVB-T re-
ceiver. Since October 2008 he is actively working with the “DSP
for Communications Lab” (DSPCOLA) of the Dept. of Infor-
mation Engineering, University of Pisa, under the supervision of
Prof. Marco Luise.
   Since January 2010 he is attending his PhD studies in radio
transmission systems at University of Pisa. His research inter-
ests are in the areas of digital communications, signal processing
and estimation theory. His current research topics focus on effi-
cient signal processing algorithms for sofware defined radio and
estimation theory applied to cognitive radio.
\end{IEEEbiographynophoto}






\end{document}